\documentclass{pasj00}
\usepackage{times}

\begin{document}
\SetRunningHead{Cao \& Xu}{The radiation properties of an
accretion disk} \Received{2002/03/29} \Accepted{2002/11/08}

\title{The radiation properties of an accretion disk with a non-zero
torque on its inner edge}

\author{Xinwu \textsc{Cao}
 }
\affil{Shanghai Astronomical Observatory, Chinese Academy of Sciences,
80 Nandan Road, Shanghai, 200030, China}

\email{cxw@center.shao.ac.cn}

\and

\author{Y.D. \textsc{Xu}}
\affil{Physics Department, Shanghai Jiaotong University, Shanghai,
200030, China}
\email{ydxu@sjtu.edu.cn}

%


\maketitle

\newcommand{\bd}{\begin{displaymath}}
\newcommand{\ed}{\end{displaymath}}
\newcommand{\be}{\begin{equation}}
\newcommand{\ee}{\end{equation}}

\begin{abstract}
The structure of the inner edge of the accretion disk around a
black hole can be altered, if the matter inside the marginally
stable orbit is magnetically connected to the disk. In this case,
a non-zero torque is exerted on its inner edge, and the accretion
efficiency $\epsilon$ can be much higher than that in the standard
accretion disk model. We explore the radiation properties of an
accretion disk at its sonic point around a black hole with a
time-steady torque exerted on the inner edge of the disk. The
local structure of the accretion flow at the sonic point is
investigated in the frame of general relativity. It is found that
the accretion flow will be optically thin at its sonic point for
most cases, if the additional accretion efficiency
$\delta\epsilon$ caused by the torque is as high as $\sim$10 \%.
The results imply that the variable torque may trigger transitions
of the flow between different accretion types.
\end{abstract}

\section{Introduction}

Most work on accretion disks around black holes are on the
assumption that there is no stress at the inner edge of the disk
(Shakura \& Sunyaev 1973). The inner edge of the disk should occur
very close to the radius of the marginally stable orbit $r_{\rm
ms}$, since the plunging matter in the region of unstable orbits
rapidly becomes causally disconnected from the disk. Recently,
Krolik (1999) questioned this assumption and  suggested that the
matter inside the marginally stable orbit can remain magnetically
connected to the disk. If this is the case, an additional torque
would be exerted on the inner edge of the disk. Gammie (1999) has
shown that the torque enhances the amount of energy released in
the disk. The energy of the matter in the plunging region is
extracted and then released in the disk. The extracted energy is
transported outwards by the stresses in the disk. The torque
alters the disk structure, and then the radiation spectrum of the
disk (Agol \& Krolik 2000, hereafter AK00). In their calculations,
AK00 assumed that the circular motion of the matter in the disk is
Keplerian, and the released gravitational energy of the accreting
matter is balanced by the radiation locally, i.e., no radial
energy advection has been considered in their calculations.

In this work, we investigate the local structure of the accretion disk at
the sonic point with a non-zero torque exerted on its inner edge.
The radiation properties of the disk at its sonic point is explored.
In our calculations, the radial energy advection in the flow
is considered, and the circular velocity of the flow is not limited
to the Keplerian value.

\section{Model}

AK00 assumed that the circular motion of the matter in the disk is
Keplerian instead of considering the radial force balance, and the inner
edge of the disk is therefore set at $r_{\rm ms}$ manually.
They further assumed that the generated energy is dissipated locally in
the disk, i.e., the radial energy advection has not been considered
in their work. The general relativistic effects are included in their
calculations, as done by Novikov \& Throne (1973).

For a transonic accretion flow around a black hole, we take the
radius of sonic point $r_{\rm s}$ as the inner edge of the disk.
We assume that a time-steady torque is exerted on the inner edge
of the disk, as done by AK00 on the assumption of a Keplerian disk
(and also see Paczynski (2000) and  Afshordi \& Paczynski (2002)).
We include the angular and radial momentum equations in our
calculations and the circular velocity of the  flow is not limited
to the Keplerian value. Our calculations are in the frame of
general relativity, and the radial energy advection in the flow is
included in the energy equation. The local structure of the flow
at the sonic point is therefore available. The torque may be
variable and extended to a region of the disk outside the radius
of sonic point. For simplicity, we restrict our calculations to
the simplest case as AK00, i.e., a time-steady torque is exerted
on the inner edge of the disk.

The magnetic field plays an important role in angular momentum
transfer of the matter in the disk. The value of viscosity
$\alpha$ can be determined by the  magnetohydrodynamical processes
taking place in the accretion flow (Hawley 2000). The value of
$\alpha$ of the flow may vary with time and radius. The angular
momentum transfer in the disk caused by any specific
magnetohydrodynamical processes may be described approximately by
the parameter $\alpha$, though the precise value of $\alpha$ and
its variation is unavailable unless the detailed
magnetohydrodynamical processes in the flow are included in the
numerical simulations. The angular momentum transfer of the matter
in the flow caused by the magnetohydrodynamical processes in the
flow may correspond to a certain value of $\alpha$, and the torque
exerted on the inner edge of the disk is governed by the same
magnetohydrodynamical processes. So, the values of $\alpha$ and
the torque are determined simultaneously  by the
magnetohydrodynamical processes taking place in the disk. In order to
avoid being involved in the complicated magnetohydrodynamical processes
taking place in the flow, we adopt $\alpha$-viscosity and another
independent parameter $\delta\nu$ to describe the angular momentum
transfer of the matter in the flow and the torque on the inner
edge of the disk, respectively. This may induce some inconsistency
in our calculations, but we can explore the general radiation
properties of the disk varying $\alpha$ for a given torque. This
will not change our main conclusions on the radiation properties
of the disk (see further discussion in Sect. 5).

For a disk with an additional torque exerted on its inner edge,
the viscosity $\nu$ at the sonic point is \be \nu(r_{\rm
s})=\nu_{0}(r_{\rm s})+\delta\nu, \label{vstress} \ee where
$\nu_{0}(r_{\rm s})$ is the viscosity at the sonic point without
an additional torque, $\delta \nu$ is caused by the additional
torque exerted on the inner edge of the disk.

The main goal of this work is to explore the local structure and
radiation properties of the disk at its sonic point. The radiation
properties are mainly governed by the radiative processes and the
energy dissipated in the flow. The $\alpha$-description on the
angular momentum transfer in the disk is  therefore sufficiently
good for present investigation, though the exact value of $\alpha$
is unknown. In this work, we focus on the local structure of the
disk and the radiation properties at the sonic point of the flow.
We assume that $\alpha$-viscosity is valid in the flow at the
sonic point. The variation of $\alpha$ along radius may not affect
our main results of the radiation properties, since we focus on
the local structure of the disk at the sonic point.

\section{Equations of accretion flows in Kerr geometry}

Many authors have included general relativistic effects in their
models of accretion disks around black holes (e.g., Novikov \& Throne
1973; Lu 1985; Abramowicz et al. 1996; Peitz \& Appl 1997; Gammie
\& Popham 1998; Popham \& Gammie 1998; Manmoto 2000; etc.). In
this work, we mainly adopt the equations presented by Abramowicz
et al. (1996, hereafter A96). One may refer to A96 for details.

The metric of a Kerr black hole on the equatorial plane takes the form
given by Novikov \& Thorne (1973) ($G=c=1$),

\be
ds^2=-{\frac {r^2\Delta}{A}}
+{\frac {A}{r^2}}(d\varphi-\omega dt)^2
+{\frac {r^2}{\Delta}}dr^2+dz^2,
\label{metric}
\ee
where
\be
\Delta=r^2-2Mr+a^2,
\label{delta}
\ee
\be
A=r^4+r^2a^2+2Mra^2,
\label{a}
\ee
and
\be
\omega={\frac {2Mar}{A}}.
\label{omega}
\ee
Here $M$ is the black hole mass and $a$ is the specific angular
momentum of the Kerr black hole.

The Lorentz factor $\gamma$ is defined by

\be
\gamma={\frac {1}{\sqrt{1-(v^{(\varphi)})^2-(v^{(r)})^2} } },
\label{gamma1}
\ee
where
\be
v^{(\varphi)}=\tilde R\tilde \Omega.
\label{vphi}
\ee
The quantities $\tilde R$ and $\tilde\Omega$ are defined by
\be
{\tilde R}^2= {\frac {A^2}{r^4\Delta}},~~~~\tilde\Omega=\Omega-\omega.
\label{trtom}
\ee

The radial velocity component $V$ is given by

\be
{\frac {V}{\sqrt{1-V^2}}}=\gamma v^{(r)}=u^r g_{rr}^{1/2}.
\label{v1}
\ee

Combining Eqs. (\ref{gamma1})-(\ref{v1}),
we have
\be
\gamma^2=\left({\frac {1}{1-{\tilde\Omega}^2{\tilde R}^2}}\right)
\left( {\frac {1}{1-V^2}}\right),
\label{gamma2}
\ee
and
\be
V={\frac {v^{(r)}} {\sqrt{1-{\tilde\Omega}^2{\tilde R}^2}} }.
\label{v2}
\ee

The angular momentum conservation is

\be
{\frac {\dot M}{2\pi r}}{\frac {dl}{dr}}
+{\frac { 1}{r}}{\frac {d}{dr}}
\left(\Sigma\nu A^{3/2} {\frac {\Delta^{1/2}\gamma^3}{r^4}}
{\frac {d\Omega}{dr}}\right)=0,
\label{angular}
\ee
where
\be
l=\gamma\left({\frac {A^{3/2}}{r^3\Delta^{1/2}}}\right)
\tilde\Omega
\label{l}
\ee
is the specific angular momentum per unit mass, and the term of the
angular momentum carried by the vertical flux of radiation has been
neglected (A96).

Integrating Eq. (\ref{angular}), we have

\be
{\frac {\dot M} {2\pi}}(l-l_0)=-\Sigma\nu A^{3/2} {\frac
{\Delta^{1/2}\gamma^3}{r^4}} {\frac {d\Omega}{dr}}, \label{intang}
\ee
where $l_0$ is the integral constant. We assume that an
additional torque is exerted on the inner edge of the disk:
$r=r_{\rm s}$, here $r_{\rm s}$ is the radius of sonic point.
At the sonic point, the viscosity $\nu(r_{\rm s})$ is described by
Eq. (\ref{vstress}). The additional accretion efficiency caused by
the torque is $\delta\epsilon$. We can obtain a relation
at $r=r_{\rm s}$:

\be {\frac {\delta\epsilon\dot M}{2\pi \Omega}} =-\Sigma\delta\nu
A^{3/2} {\frac {\Delta^{1/2}\gamma^3}{r_{\rm s}^4}} \left. {\frac
{d\Omega}{dr}} \right |_{r=r_{\rm s}}. \label{torque} \ee In the
absence of an additional torque, the viscosity $\nu_{0}(r_{\rm
s})$ at the sonic point is very low, namely zero-torque
approximation. In this work, we consider the cases with
significant additional torques on the inner edges of the disks, so
$\delta\nu\gg \nu_{0}$ is usually satisfied at the sonic point.
For simplicity, we use an approximation: $\nu(r_{\rm s})\simeq
\delta\nu$ in our calculations. We further assume that
$\alpha$-viscosity  is valid even at the sonic point. The value of
$\alpha$ at the sonic point may be different from that in the
region outside the radius of sonic point.

The radial component of the momentum conservation is
\be
{\frac {V}{1-V^2}}{\frac {dV}{dr}}
={\frac {\cal A}{r}}
-{\frac {1}{\Sigma}}{\frac {dP}{dr}},
\label{radial}
\ee
where
\be
{\cal A}=-
{\frac {MA}{r^3\Delta\Omega_{\rm K}^+\Omega_{\rm K}^-}}
{\frac {(\Omega-\Omega_{\rm K}^+)(\Omega-\Omega_{\rm K}^-)}
{1-{\tilde\Omega}^2{\tilde R}^2}},
\ee
and $P=2Hp$ is the vertical integrated pressure.

The angular velocities of the corotating and counterrotating Keplerian
orbits are
\be
\Omega_{\rm K}^{\pm}=\pm
{\frac {M^{1/2}}{r^{3/2}\pm aM^{1/2}}}.
\ee

The surface energy generation rate $F^+$ and the cooling rate
$F^-$ are \be F^+=\nu\Sigma{\frac {A^2}{r^6}}\gamma^4 \left({\frac
{d\Omega}{dr}}\right)^2, \label{fplus} \ee and \be F^-={\frac
{16\sigma T^4}{3\tau}}, \label{fminus} \ee respectively.  Here we
only consider the optically thick case.

The energy equation can be written as
\be
F^{\rm adv}=F^+-F^-.
\label{energy}
\ee

The equation of state is \be p={\frac {1}{3}}aT^4+{\frac {\rho
kT}{\mu m_{\rm H}}}, \label{state} \ee where the ratio of the
specific heats of the gas $\gamma_{\rm g}=5/3$ has been adopted.
The magnetic pressure has been neglected, since we restrict our
calculations to the optically thick case.  For the optically thick
case, the radiation processes are not relevant to the magnetic
pressure.

Using Eq. (\ref{state}), the entropy gradient can be calculated by
\be
TdS={\frac {p}{\rho}}
\left[\left(12-{\frac {21}{2}}\beta\right)d\ln T-(4-3\beta)d\ln\rho
\right],
\label{entropy}
\ee
where $\beta=p_{\rm g}/p$.

The advection cooling rate due to the radial motion of the gas is
\bd
F^{\rm adv}=-{\frac {\dot M}{2\pi r}}T{\frac {dS}{dr}}
\ed
\be
=-{\frac {\dot M}{2\pi r}}
{\frac {p}{\rho}}
\left[\left(12-{\frac {21}{2}}\beta\right){\frac {d\ln T}{dr}}
-(4-3\beta){\frac {d\ln\rho}{dr}}\right].
\label{fadv}
\ee

The vertical force balance gives
\be
{\frac {p}{\rho H^2}}
=\gamma^2{\frac {M}{r^3}}
\left[
{\frac {(r^2+a^2)^2+2\Delta a^2}
{(r^2+a^2)^2-\Delta a^2} } \right ].
\label{vertical}
\ee

Substituting Eqs. (\ref{state}), (\ref{fadv}) and  (\ref{vertical})
into Eq. (\ref{radial}), we can rewrite the radial motion equation as
\be
{\frac {dV}{dr}}={\frac {N}{D}}(1-V^2),
\label{radialf}
\ee
where
\be
D=V-{\frac {72-51\beta-9\beta^2}
{56-45\beta-3\beta^2}}
{\frac {c_{\rm s}^2}{V}},
\ee
and
\bd
N={\frac {\cal A}{r}}
+{\frac {72-51\beta-9\beta^2}
{56-45\beta-3\beta^2}}c_{\rm s}^2
{\frac {d}{dr}}\ln \Delta^{1/2}
\ed
\bd
+{\frac {-8+3\beta+3\beta^2}
{56-45\beta-3\beta^2}}c_{\rm s}^2{\frac {d}{dr}}
\ln \left \{
\gamma^2{\frac {M}{r^3}}
\left[
{\frac {(r^2+a^2)^2+2\Delta a^2}
{(r^2+a^2)^2-\Delta a^2} } \right ]\right\}
\ed
\be
+{\frac {16-12\beta}{56-45\beta-3\beta^2}}
{\frac {2\pi rF^{\rm adv}}{\dot M}}.
\label{n}
\ee

The sonic point is defined by the condition \be D=N=0,
\label{sonic} \ee and we have the radial velocity of the flow  at
the sonic point: \be V_{\rm s}=\left( {\frac {72-51\beta-9\beta^2}
{56-45\beta-3\beta^2}}  \right)^{1/2} c_{\rm s}. \ee

The standard $\alpha$-viscosity
\be
\nu={\frac {2}{3}}\alpha c_{\rm s} H,
\ee
is adopted in this work.

Substituting Eqs. (\ref{torque}), (\ref{fplus}), (\ref{fminus})
and (\ref{fadv}) into Eqs. (\ref{n}) and (\ref{sonic}), the
physical quantities $\rho_{\rm s}$ and $T_{\rm s}$ of the flow at
the sonic point are available, if the parameters $M$, $\dot M$,
$a$, $r_{\rm s}$, $\alpha$, $\delta\epsilon$ and $\Omega_{\rm s}$
are specified.

The effective optical depth in the vertical direction is
\be
\tau_{\rm eff}=0.5 \Sigma\sqrt{(\kappa_{\rm es}+\kappa_{\rm ff})
\kappa_{\rm ff}},
\label{taueff}
\ee
where $\kappa_{\rm es}$ and $\kappa_{\rm ff}$ are the electron
scattering opacity and free-free opacity, respectively.

\begin{figure}
  \begin{center}
    \FigureFile(90mm,90mm){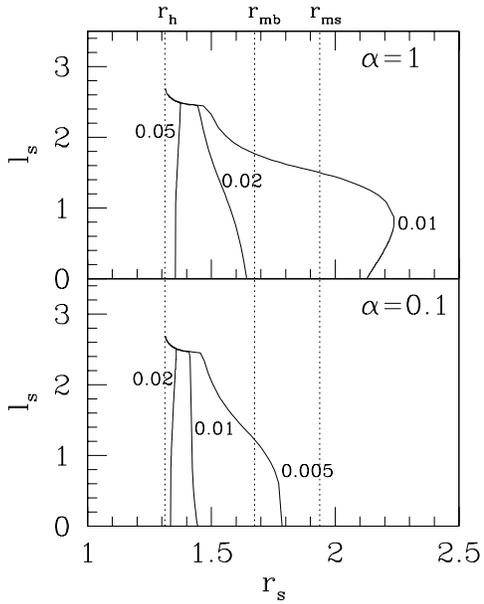}
  \end{center}
  \caption{
The parameter plane $r_{\rm s}-l_{\rm s}$ for all possible
solutions with different values of $\delta\epsilon$ (labeled near
the curves). The parameters: $M=10^9 M_{\odot}$, $a=0.95$ and
$\dot M/{\dot M}_{\rm Edd}=0.001$ are adopted. The parameters in
the left side of the curves are for optically thick solutions,
i.e., $\tau_{\rm eff}>1$. The radius of the black hole horizon
$r_{\rm h}$, radius of the minimal bound circular orbit $r_{\rm
mb}$ and radius of the marginally stable orbits $r_{\rm ms}$ are
marked in the figure. The upper panel is for $\alpha=1$, while the
lower panel is for $\alpha=0.1$.
  }
\end{figure}

\begin{figure}
  \begin{center}
    \FigureFile(90mm,90mm){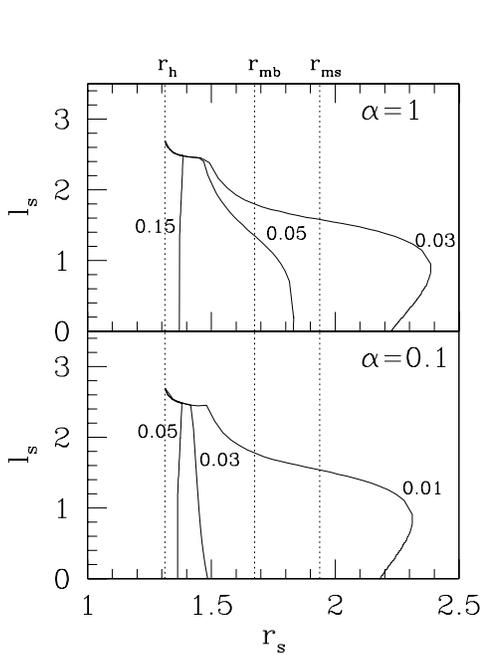}
  \end{center}
  \caption{
The same as Fig. 1, but for
$\dot M/{\dot M}_{\rm Edd}=0.1$.
  }
\end{figure}

\begin{figure}
  \begin{center}
    \FigureFile(90mm,90mm){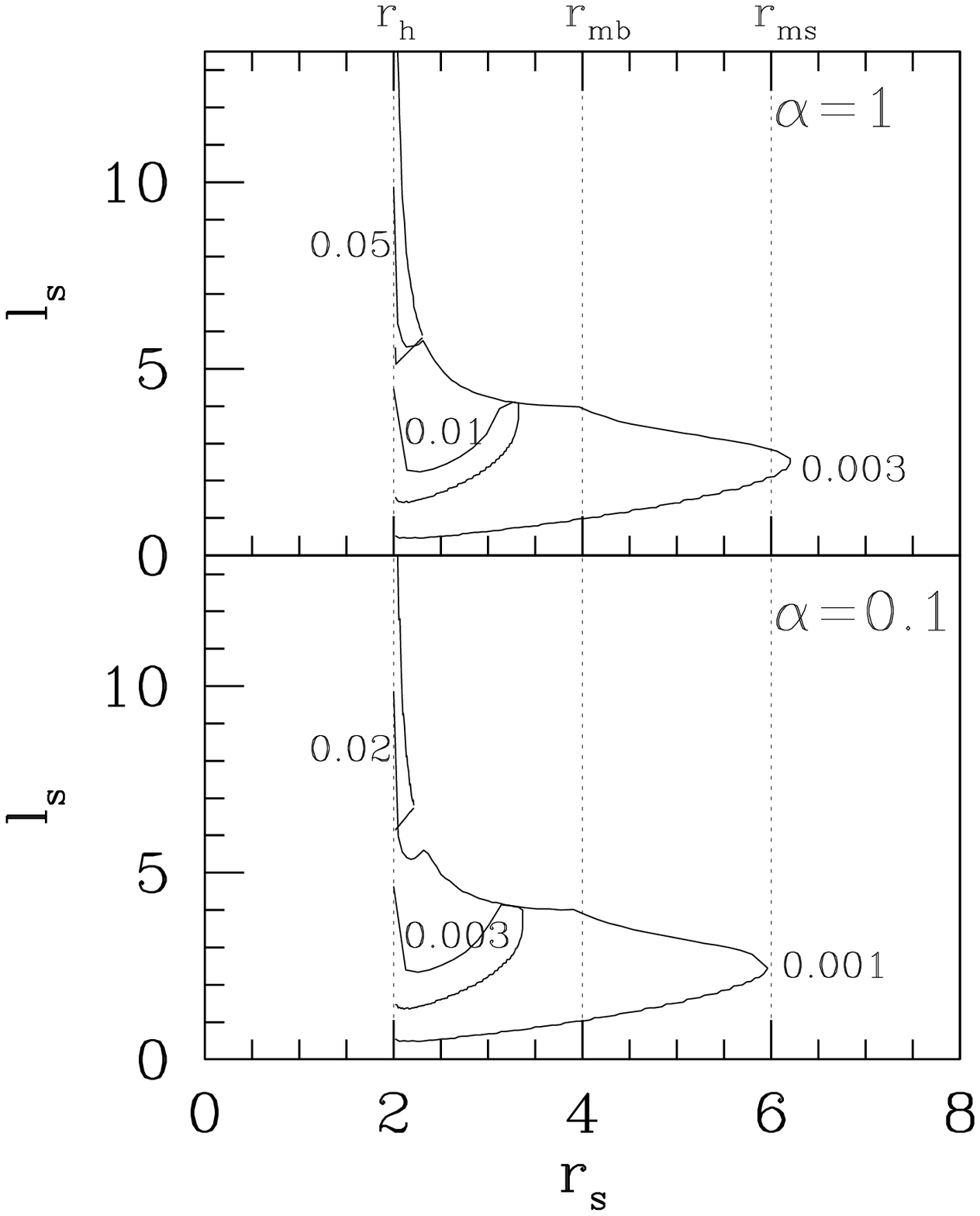}
  \end{center}

  \caption{
The same as Fig. 1, but for a Schwarzschild black hole ($a=0$).
  }
\end{figure}

\begin{figure}
  \begin{center}
    \FigureFile(90mm,90mm){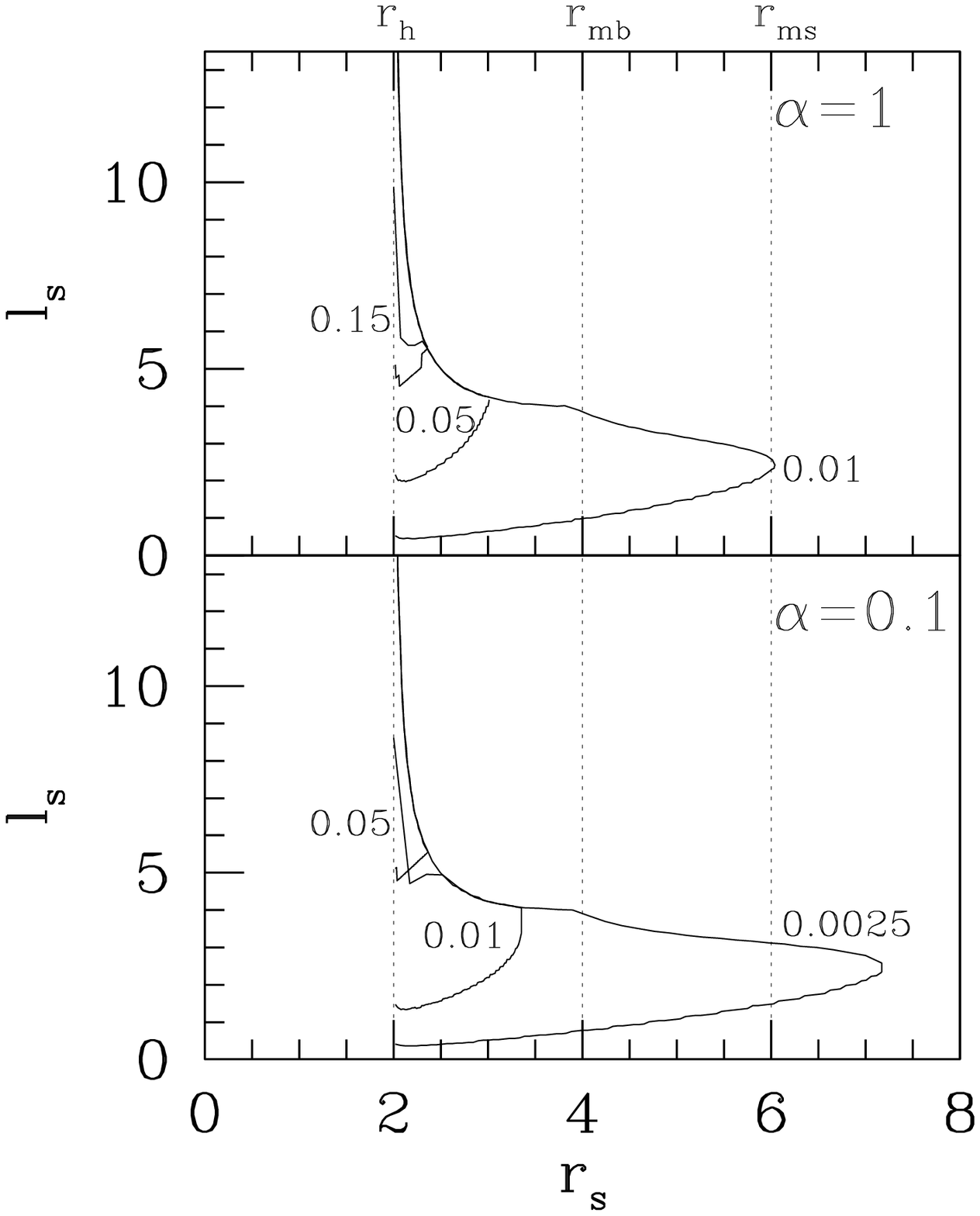}
  \end{center}

  \caption{
The same as Fig. 2, but for a Schwarzschild black hole ($a=0$).
  }
\end{figure}

\begin{figure}
  \begin{center}
    \FigureFile(90mm,90mm){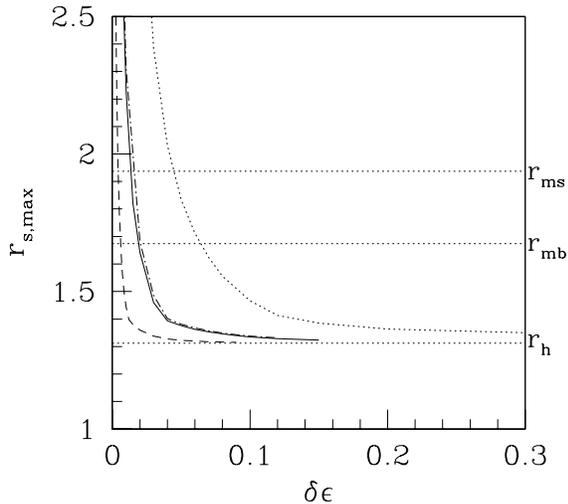}
  \end{center}
  \caption{
The maximal radius of sonic point $r_{\rm s,max}$ as a function of
$\delta\epsilon$ for $a=0.95$. The different values of the
parameters are adopted: $\dot M/{\dot M}_{\rm Edd}=0.001$,
$\alpha=1$ (solid line); $\dot M/{\dot M}_{\rm Edd}=0.001$,
$\alpha=0.1$ (dashed); $\dot M/{\dot M}_{\rm Edd}=0.1$, $\alpha=1$
(dotted) and $\dot M/{\dot M}_{\rm Edd}=0.1$, $\alpha=0.1$
(dot-dashed).
 }
\end{figure}

\begin{figure}
  \begin{center}
    \FigureFile(90mm,90mm){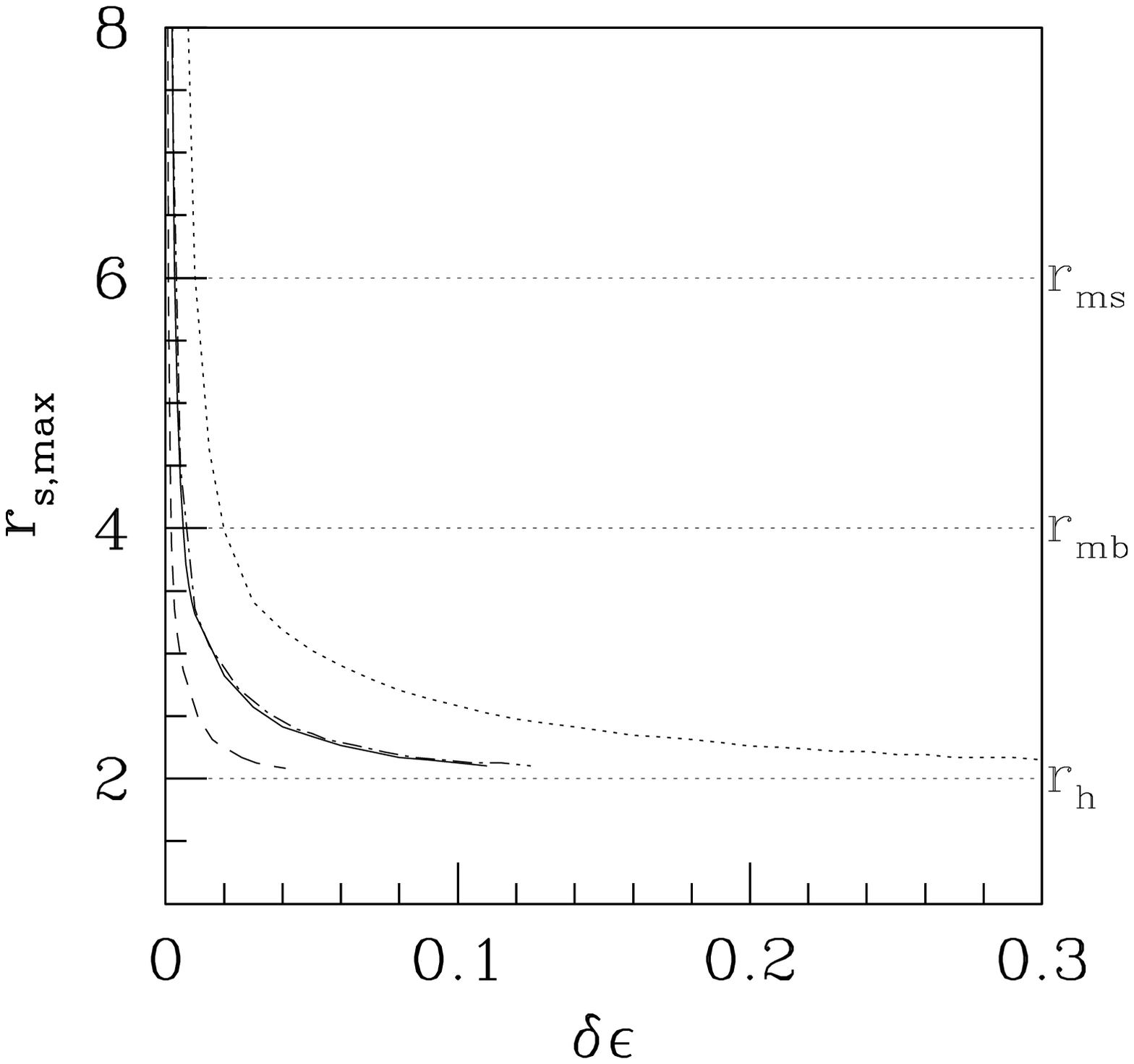}
  \end{center}
  \caption{
The same as Fig. 5, but for a Schwarzschild black hole
($a=0$).
 }
\end{figure}

\section{Results}

In principle, the set of equations describing the accretion flow
around a black hole can be integrated with suitable outer boundary
conditions. The solution is required to satisfy the regularity
condition (\ref{sonic}) at the sonic point $r=r_{\rm s}$ for a
given $\delta\epsilon$. The specific angular momentum $l_{\rm s}$
at the sonic point and the radius of sonic point $r_{\rm s}$ are
simultaneously given by the numerical solution. In the case of
$\delta\epsilon=0$, $l_{\rm s}\simeq l_0$. The main numerical
difficulties in solving the set of equations would be the
particular value of $l_{\rm s}$ which should satisfy the
regularity condition (\ref{sonic})(see A96 for the cases of
$l_{\rm s}\simeq l_0$).

In this work, we search all possible solutions in the parameter
plane $r_{\rm s}-l_{\rm s}$ instead of integrating the equations
with outer boundary conditions numerically. Using this method, we
can study the local properties of the disk at the sonic point. All
possible solutions satisfying condition (\ref{sonic}) can be
found. The advantage of this approach is that all physical
solutions satisfying both the outer boundary conditions and the
inner regularity condition (\ref{sonic}) will be included in the
solutions found in this way, though some solutions we find would
be unphysical.

As discussed in Sect. 3, we only consider the optically thick case
(see Eq. (\ref{fminus})). So, all optically thick solutions
should satisfy $\tau_{\rm eff}>1$ at the sonic point.

In this work, we focus on the geometrically thin accretion disk,
all calculations are therefore  performed for accretion rates
$\dot{M}/\dot{M}_{\rm Edd}\leq 0.1$. We found that all possible
optically thick disk solutions obtained in this work
do satisfy the thin disk approximation $H/r\le 0.1$ at the sonic point.
The parameter planes $r_{\rm
s}-l_{\rm s}$ for all possible optically thick solutions are
present in Figs. 1-4, for different values of
$\dot{M}/\dot{M}_{\rm Edd}$, $\alpha$, and $a$. We found that the
parameter plane $r_{\rm s}-l_{\rm s}$ for all possible optically
thick solutions is sensitive to these parameters. It is
interesting to find that there is an upper limits on the radius of
sonic point in the solution planes $r_{\rm s}-l_{\rm s}$ for
given parameters. The upper limits on the radius of  sonic point
$r_{\rm s,max}$ for possible solutions are plotted in Figs. 5 and
6.

\section{Discussion of the results}

The physical, global solutions for optically thick flows are
available by integrating a set of equations with suitable inner
and outer boundary conditions. The problem is that the search for
such a parameter plane for all physical global solutions with all
possible different inner and outer boundary conditions would be
very difficult. The parameter plane $r_{\rm s}-l_{\rm s}$ for
physical, global solutions will be included in the present
parameter plane $r_{\rm s}-l_{\rm s}$, but the global solution
will set a more strict constraint on the existence of optically
thick flows. So, the conclusion is tight that no physical,
optically thick disk solution at the sonic point will be present
outside the parameter plane $r_{\rm s}-l_{\rm s}$ obtained here,
though the possibility cannot be ruled out that the flow is
optically thin at the sonic point while it becomes optically thick
at a large radius outside the sonic point.

It is found that the location of the sonic point for an optically
thick accretion flow moves towards the horizon of the hole with
the increase of $\delta\epsilon$. For high $\delta\epsilon$, no
optically thick accretion flow  at the sonic point is present. A
higher $\delta\epsilon$ means that much energy of the plunging
matter inside the marginally stable orbit is extracted to the
disk, and the disk is heated to a higher temperature. The
free-free opacity $\kappa_{\rm ff}$ decreases with the increase of
 temperature. The disk then becomes optically thin.

Comparing the results for different values of $\alpha$, we find
that the accretion flow with a low $\alpha$ will be optically thin
even for a low $\delta\epsilon$, i.e., the optically thick flows
at the sonic point with low $\alpha$ are present only for low
$\delta\epsilon$ cases. A high $\alpha$ makes the energy extracted
from the plunging matter inside the inner edge of the disk
transported outwards efficiently. Therefore, only a small fraction
of the energy is released locally at the inner edge of the disk,
and the flow can still remain optically thick for a relatively
high $\delta\epsilon$. For the lower $\alpha$ case, much energy is
released locally and the disk is heated to a higher temperature,
and the disk becomes optically thin at the sonic point. For a high
accretion rate ${\dot m}={\dot M}/{\dot M} _{\rm Edd}$, the
surface density of the disk is so high that the disk is optically thick
even for a relatively high $\delta\epsilon$ (compare results in
Figs 1-4 for different values of $\dot m$).

The sonic point of the accretion flow usually locates at the
radius close to the marginally stable orbit, since the matter will
plunge to the hole rapidly inside $r_{\rm ms}$. The radius of
 sonic point is in principle available by integrating the set of
the equations with suitable outer and inner boundary conditions.
In the present work, we only consider the inner boundary condition at
the sonic point, and only the range of the radius of sonic point
is given. From Figs. 5 and 6, we find that $\delta\epsilon\le
0.07$ should be satisfied for an optically thick flow at the sonic
point around a Kerr black hole ($a=0.95$), if $r_{\rm s}\ge r_{\rm
mb}$ is assumed (see Fig. 5). For a Schwarzschild black hole,
$\delta\epsilon\le 0.02$ should be satisfied (see Fig. 6). In our
calculations, we found that the results are insensitive to the
black hole mass.

Recently, the {\it XMM-Newton} observation of the Seyfert 1 galaxy
MCG$-$6-30-15 reveals an extremely broad and red-shifted Fe
K$\alpha$ line (Wilms et al. 2001). The explanation on the
observed spectrum requires the disk to have a very steep emissivity
profile with an index around 4.3-5.0. A possible explanation is
provided by the magnetic connection between the inner region
of the disk and the plunging matter inside the inner edge of the
disk (Wilms 2001; Krolik 1999; AK00). Another possible explanation
is based on the magnetic connection between a rotating black
hole and a disk (Wilms 2001; Li 2002a,b). However, it is also
suggested that the origin of the broad Fe K$\alpha$ line can be
explained in the frame of an illuminated relativistic accretion
disk (Martocchia, Matt \& Karas 2002). AK00 suggested that the
locally generated surface energy flux scales as $r^{-7/2}$ at
large $r$ in the limit of infinite efficiency or zero accretion
rate in their model, rather than that scales as $r^{-3}$ in the
standard thin disk model. This is helpful to explain the required
steep emissivity profile. In this work, we have investigated the
local structure and radiation properties of the disk at the sonic
point. The global solution to this problem could be tested against
the observed broad fluorescent Fe K$\alpha$ line in MCG$-$6-30-15,
which will be given in our future work.

In this work, we find that the radiation properties of the disk at
the sonic point are sensitive to the torque exerted on the inner
edge of the disk. The torque may probably be variable (e.g., Armitage,
Reynold \& Chiang 2001; Hawley 2001; Hawley \& Krolik 2001, 2002;
Reynold \& Armitage 2001; etc.). Our results imply that the
variable torque may trigger transitions of the flow between
optically thick and optically thin accretion types. The global
solutions are necessary to attack this problem, which is beyond
the scope of the present work,  and it will be reported in our
future work.

In present calculations, we have assumed that the torque is
exerted on the inner edge of the disk, as done by AK00. If the
torque is exerted on the extended region outside the inner edge of
the disk, the disk can be optically thick at the sonic point
even for a relatively higher $\delta\epsilon$ than that reported here.

\section*{Acknowledgments}
XC is grateful to J.F. Lu for helpful discussion. We thank the anonymous
referee for his/her helpful comments.  This work is supported by
the NSFC (No. 10173016) and NKBRSF (No. G1999075403).

{}

\end{document}